\documentclass[final,3p,times]{elsarticle}

\usepackage{amssymb}
\usepackage{float}
\usepackage{hyperref}
\hypersetup{pdftex,colorlinks=true,allcolors=blue}
\usepackage{hypcap}
\usepackage{color}

\journal{Physica E}

\begin{document}

\begin{frontmatter}

\title{Semimetal behavior of bilayer stanene}

\author[label1]{I. Evazzade\corref{cor1}}
\cortext[cor1]{Corresponding auther}
\ead{i.evazzade@mail.um.ac.ir}
\author[label1]{M. R. Roknabadi}
\author[label2]{T. Morshedloo}
\author[label1]{M. Modarresi}
\author[label3,label4]{Y. Mogulkoc}
\author[label1]{H. Nematifar}

\address[label1]{Department of Physics, Faculty of Science, Ferdowsi University of Mashhad, Mashhad, Iran}
\address[label2]{Nano Structured Coatings Institute of Yazd Payame Noor University, 89431-74559, Yazd, Iran}
\address[label3]{Department of Engineering Physics, Ankara University, 06100, Tandogan, Ankara, Turkey}
\address[label4]{Institute for Molecules and Materials, Radboud University, Heyendaalseweg 135, 6525AJ Nijmegen,The Netherlands}

\begin{abstract}
Stanene is a two-dimensional (2D) buckled honeycomb structure which has been studied recently owing to its promising electronic properties for potential electronic and spintronic applications in nanodevices. In this article we present a first-principles study of electronic properties of fluorinated bilayer stanene. The effect of tensile strain, intrinsic spin-orbit and van der Waals interactions are considered within the framework of density functional theory. The electronic band structure shows a very small overlap between valence and conduction bands at the $\Gamma$ point which is a characteristic of semimetal in fluorinated bilayer stanene. A relatively high value of tensile strain is needed to open an energy band gap in the electronic band structure and the parity analysis reveals that the strained nanostructure is a trivial insulator. According to our results, despite the monolayer fluorinated stanene,
the bilayer one is not an appropriate candidate for topological insulator.  
\end{abstract}

\begin{keyword}
bilayer stanene \sep semimetal behavior \sep topological insulators \sep  first-principles calculations

\end{keyword}

\end{frontmatter}

\section{Introduction}
Recently theoretical investigation of materials based on group IV elements attract considerable attention \cite{Kane2005, CWZhang2012, CWZhang2012_2, CWZhang2016, CWZhang2016_2}. Among them, Stanene the two-dimensional (2D) tin as a topological insulators has been studied thoroughly because of being ideal candidates to quantum spin Hall (QSH) effect \cite{Hasan2010, Qi2011, Yan2012, Moore2013, Rasche2013, Nigam2015}. The topological state was observed experimentally in $HgTe$ quantum wells \cite{König2007}, three dimensional structures e.g., $Bi_{1-x}Sb_{x}$ \cite{Hsieh2009}, $SnTe$ \cite{Tanaka2012}, $Bi_{2}Te_{3}$ \cite{Chen2009}, $TiBiSe_{2}$ \cite{Kuroda2012} and predicted theoretically for two dimensional nanostructures \cite{Chou2014, Cao2015, R.Zhang2015, Ren2015, Kane2005, Liu2011, H.Zhang2015}. These materials are associated with topological invariant which is different from ordinary insulators having topologically protected gapless edge states that are robust against time-reversal-invariant weak disorders \cite{Kane2005, Wang2012}. In spite of the fact that 2D materials have been on the focus of a large number of scientific researches, not all of them are fully utilized as QSH insulators. Graphene has been predicted to be a QSH insulator by Kane and Mele at low temperatures with a too small energy gap ($\approx$ 3-10 meV) which is opened by spin-orbit coupling at the Dirac point \cite{Kane2005}. The functionality of QSH insulators is having an experimentally accessible gap at room temperature that is due to large spin-orbit coupling strength \cite{Xu2013}. \par

In the pursuit of large-gap QSH insulators, many novel structures have been predicted \cite{Nigam2015, Chou2014, Cao2015, R.Zhang2015, Xu2013, CWZhang2016,CWZhang2016_2} including slightly buckled honeycomb lattice of tin atoms\cite{CCLiu2011}, also known as stanene, which recently has been fabricated by molecular beam epitaxy in laboratory \cite{Zhu2015}. Due to the relatively large atomic mass of tin atoms the intrinsic spin-orbit interaction opens an energy gap in the electronic structure of stanene which is around 70 meV \cite{Xu2013, Modarresi2015}. Theoretical calculations predict a larger gap for halogenated monolayer \cite{Xu2013} and ethynyle-derivative \cite{CWZhang2016} stanene, as well as chemically decorated plumbene \cite{CWZhang2016_2}, that suggest these structures for application in nanodevices based on topological insulators. Indeed, it was shown that oxygen adsorption may cause topological behavior to trivial phase transition in stanene nanoribbons \cite{MoMo2016}.\par 

From a realistic point of view, the synthesis of an exact monolayer of 2D structure on a substrate is an experimental challenge for experimental scientists. In the case of stanene, the latest attempt by molecular beam epitaxy leads to few layer stanene \cite{Zhu2015}. The main question arises from the effect of interaction between different layers of stanene on the topological phase of the nanostructure. Based on the aforementioned motivation, we investigate the effect of fluorine absorption on the electronic and topological phase of bilayer stanene. The fluorinated bilayer stanene (FBSn) is studied by using density function theory (DFT) with considering the van der Waals (vdW) interaction between different layers of structure. Owing to the heavy atomic mass of tin atoms, fully relativistic pseudopotential together with spin-orbit coupling is considered for DFT calculations. 

\section{Model and Method}
For our calculations we have used the first principles density functional calculations as implemented in the Quantum Espresso code \cite{Giannozzi2009}. The electronic calculations have been performed within the Generalized Gradient Approximation (GGA) and the Perdew-Burke-Ernzerhof (PBE) \cite{Perdew2010} exchange correlation is adopted. The semi-empirical dispersion interactions (DFT-D) model \cite{Grimme2010,Grimme2006} has been employed in order to consider the van der Waals interaction between adjacent layers of bilayer stanene. To avoid interaction between stanene layers in neighbor unit-cells, a vacuum layer $15\mathring{A}$ is set perpendicular to the stanene plane. A $21\mathord{\times}21\mathord{\times}1$ Monkhorst and Pack \cite{Monkhorst1976} k-point mesh is adopted to sample the Brillouin zone and the cut-off energy for plane wave expansion is set to $700 Ry$ in all calculations.\par
\begin{figure}
\centering{\includegraphics*[scale=0.5]{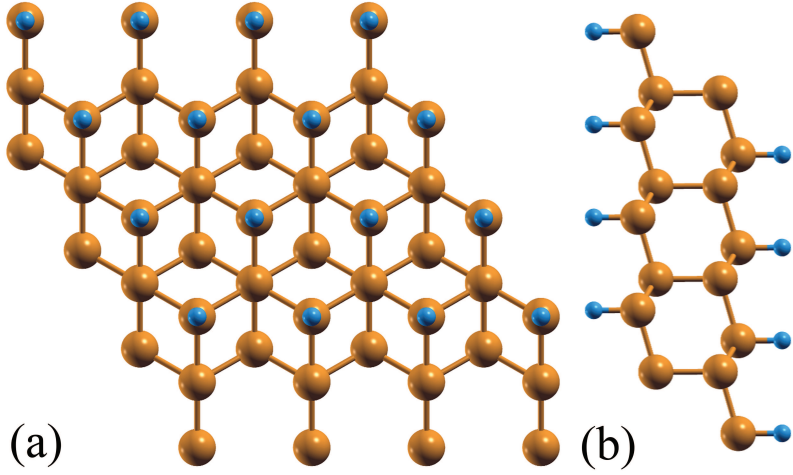}}
\caption{\label{Fig_1}(a) Top and (b) side view of atomic structure of FBSn}
\end{figure}

 
Throughout geometry optimization, cell parameters as well as atomic positions are fully relaxed until the force on atoms is less than $0.01 eV/\mathring{A}$. We consider the hexagonal unit cell for bilayer fluorinated stanene as shown in Figure \ref {Fig_1}. The side view represents the buckling structure of tin atoms. The relaxation process does not break the space inversion symmetry of bilayer fluorinated stanene. The topological invariant determines the topological/trivial phase of structure. According to the Fu-Kane work \cite{Fu2007} in a crystal with space inversion symmetry one can obtain the topological invariant by using the parity of occupied electronic bands of Time Reversal Invariant Momenta (TRIM) points \cite{RWZhang2016, YWang2016}. In the hexagonal structure, there are 4 TRIM points (1 $\Gamma$ and 3 $M$ points) in the first Brillouin zone \cite{Chou2014,R.Zhang2015}.

\section{Results}
We consider the AB stacked of bilayer stanene. Figure \ref{Fig_1} represents the top and side view of buckled structure of FBSn. Both figures are plotted using XCrySDen program \cite{Kokalj1999}. The lattice parameter of hexagonal unit cell of FBSn is $4.75\mathring{A}$ which is smaller than iodinated stanene \cite{Chou2014}. In the FBSn the buckling parameter of layers is $0.8\mathring{A}$ which is around pure stanene and the distance between adjacent layers is about $3.6\mathring{A}$ which is in quite agreement with recent epitaxial growth of stanene \cite{Zhu2015}.\par

\begin{figure}
\centering{\includegraphics*[scale=0.4]{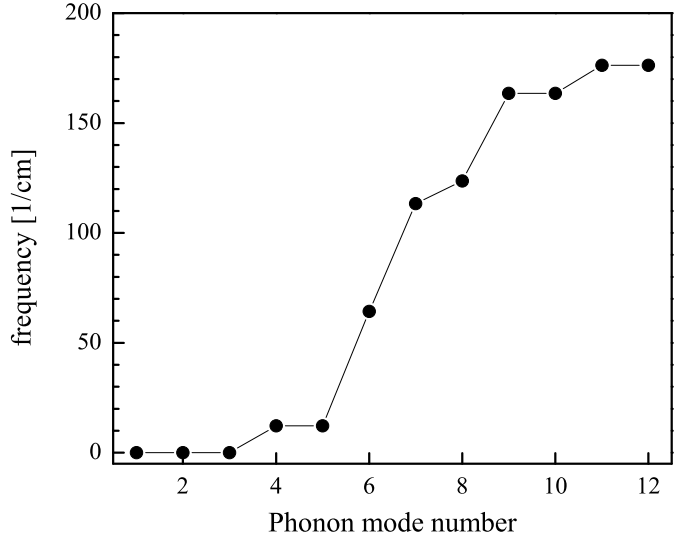}}
\caption{\label{Fig_2}Phonon frequency of bilayer stanene at the $\Gamma$ point.}
\end{figure}

The positive phonon spectrum confirms the thermal stability of material. In Figure \ref{Fig_2} the phonon frequencies of bilayer stanene at the $\Gamma$ point are plotted for all acoustic and optical branches. The maximum phonon frequency at the $\Gamma$ point is 176 $cm^{-1}$ which is almost unchanged with respect to monolayer stanene \cite{SZaveh2016}. For the monolayer stanene there is a phonon frequency gap between 50 $cm^{-1}$ and 100 $cm^{-1}$ \cite{SZaveh2016}. In the case of bilayer stanene the frequency gap is filled with one phonon state at 60 $cm^{-1}$ for the $\Gamma$ point.

\begin{figure}[!h]
\centering{\includegraphics*[scale=0.5]{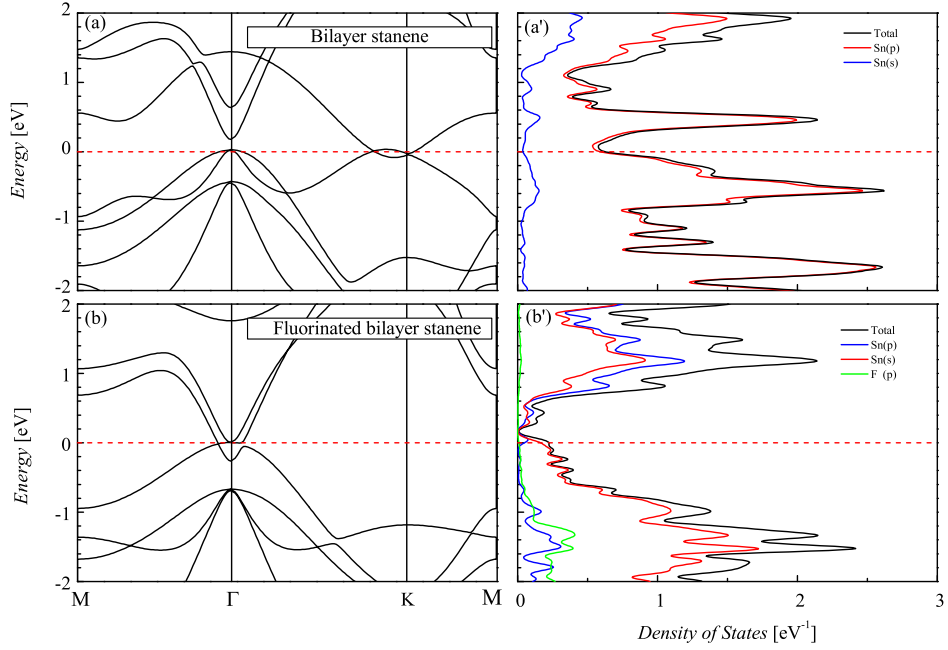}}
\caption{\label{Fig_3}Band structure and PDOS of  $(a, a')$bilayer stanene and $(b, b')$ FBSn in the absence of spin-orbit interaction.}
\end{figure}

Figure \ref{Fig_3} shows the band structure and partial density of state (PDOS) for pure bilayer stanene and FBSn in the absence of spin-orbit coupling (SOC). The band structure is plotted along the high symmetry points of the first Brillouin zone.\par
In the pure bilayer structure, the bands at the $K$ point cross the Fermi level which indicate the semimetal behaviour. In both valence and conduction bands the main contribution comes from atomic p orbitals of tin atoms.  In the FBSn structure, unlike graphene the energy valence and conduction bands at the $K$ point are separated by a relatively large energy gap. The minimum of conduction band at the $\Gamma$ point cross the Fermi level and electronic gap is zero. The PDOS indicates that the eigenstates related to the p atomic orbital of tin atoms fill the band gap at the $\Gamma$ point and FBSn becomes a semimetal. Around the Fermi level, electronic states in the valence band mostly arise from the p atomic orbitals but for conduction band the s atomic orbitals contribute more effectively. The close energies as well as difference in parities for s and p atomic orbitals allows the possibility of band inversion and the topological phase in tin 2D structures. The presence of s atomic orbitals in the conduction band shows the failure of any single orbital model for describing FBSn. The s and p atomic orbitals of fluorine atoms are far from the Fermi level and do not contribute in the electronic structure of bilayer stanene effectively. In both structures, the contribution of d atomic orbitals of tin atoms is negligible and ignored in figures.\par
\begin{figure}
\centering{\includegraphics*[scale=0.4]{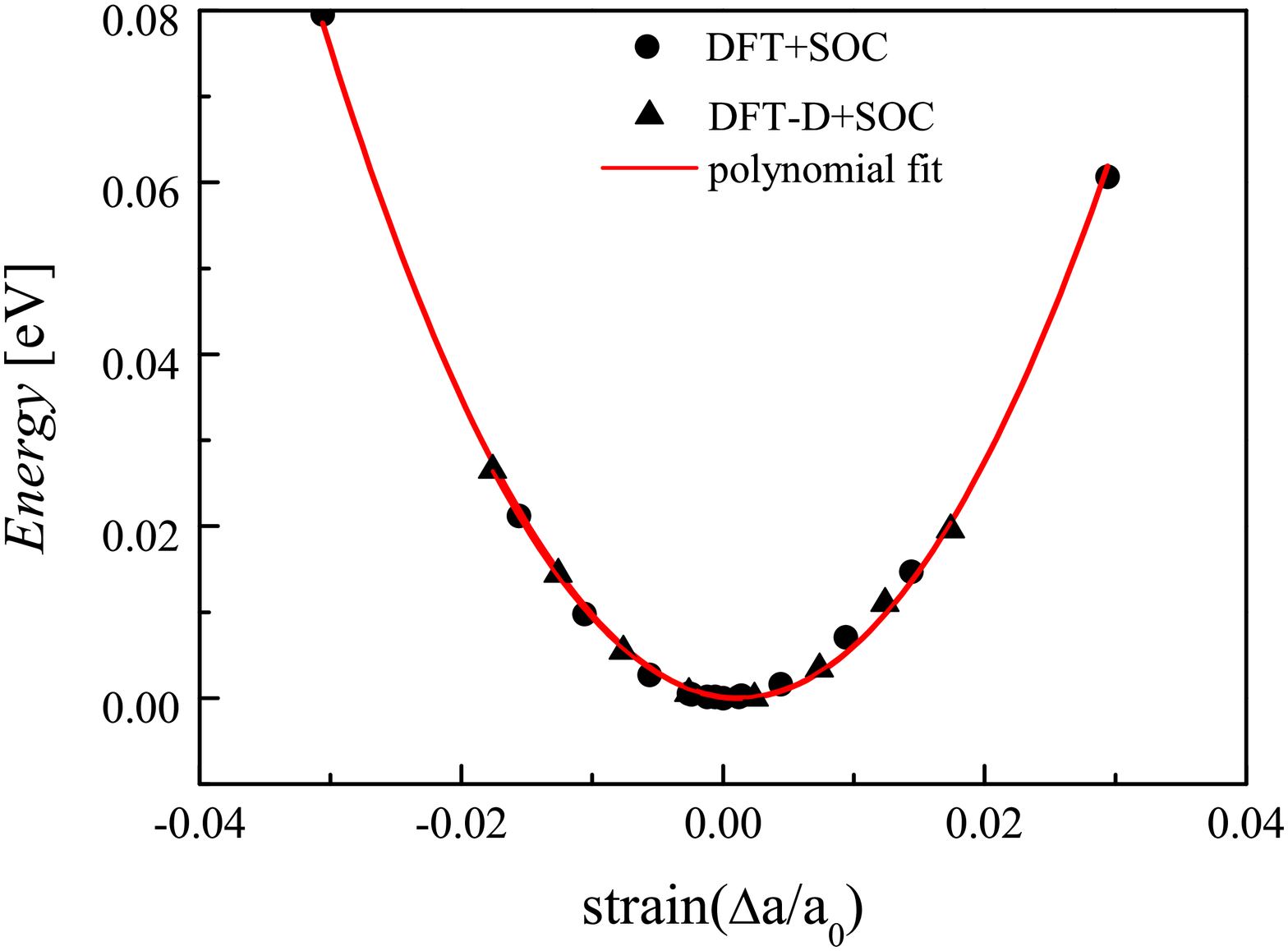}}
\caption{\label{Fig_4}Energy difference from the ground state for bilayer FBSn in the presence of spin-orbit and vdW interactions}
\end{figure}
\begin{figure}[!h]
\centering{\includegraphics*[scale=0.4]{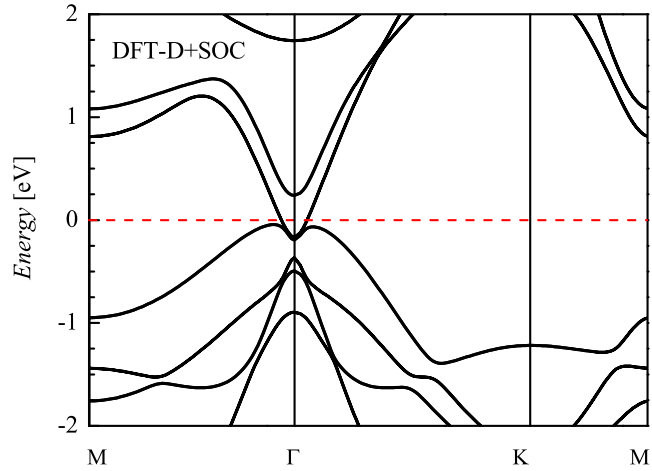}}
\caption{\label{Fig_5}Electronic band structure of FBSn with spin-orbit and van der Waals interactions}
\end{figure}

As follows, we consider the spin-orbit coupling and the van der Waals interaction between different layers of atoms in the DFT-D model. The atomic positions of structure are fully relaxed for a certain value of strain and the variation of total energy with lattice parameter is plotted in Figure \ref{Fig_4} for two different models.The polynomial curve is well fitted to the DFT and DFT-D results are shown in Figure \ref{Fig_4}. For a 2D structure one cannot define the Young's modulus due to the very large value of c axis in the hexagonal super cell. We calculate the in-plane stiffness $Y_{s}=(1/A_{0})(\partial^{2} E/\partial \varepsilon^{2})$ where $A_{0}$ is the equilibrium area of the unit cell. We obtain the $Y_{s}=60 N/m$ which is much smaller than graphene's in-plane stiffness \cite{Zhu2015, Mirnezhad2012, Peng2013} and is about twice of pure single layer stanene \cite{Modarresi2015} 
The binding energy between two fluorinated layers which is expressed as in equation \ref{eq:1},
\begin{equation}
E_{Binding}=E_{Bilayer fluorinated}-E_{Monolayer fluorinated} \label{eq:1}
\end{equation}
The binding energy between two layers is about $1.85$  $eV/unit cell$ which is increased 20\% and reaches to $2.25$ $eV/unit cell$ by considering the van der Waals interaction between two tin layers. The electronic structure of relaxed FBSn in the DFT-D model with spin-orbit interaction is shown in Figure \ref{Fig_5}.\par
In the presence of spin-orbit interaction some degeneracies in the band structure are lifted. By considering the intrinsic spin-orbit interaction, the required gap for the topological insulators does not appear in the electronic structure and FBSn still remains semimetal. For the relaxed structure the electronic band structure of system is gapless and therefore the topological invariant is ill-defined due to the mixture of valence and conduction bands at the $\Gamma$ point. The small value of in-plane stiffness ($Y_{s}$) and flexibility due to the buckling\cite{H.Zhang2015} suggest the tuning of electronic structure by applying external strain. For a structure with lattice parameter $a$ , the applied strain is defined as $\varepsilon=(a-a_{\circ})/a_{\circ}$. \par
\begin{figure}[!h]
\centering{\includegraphics*[scale=0.4]{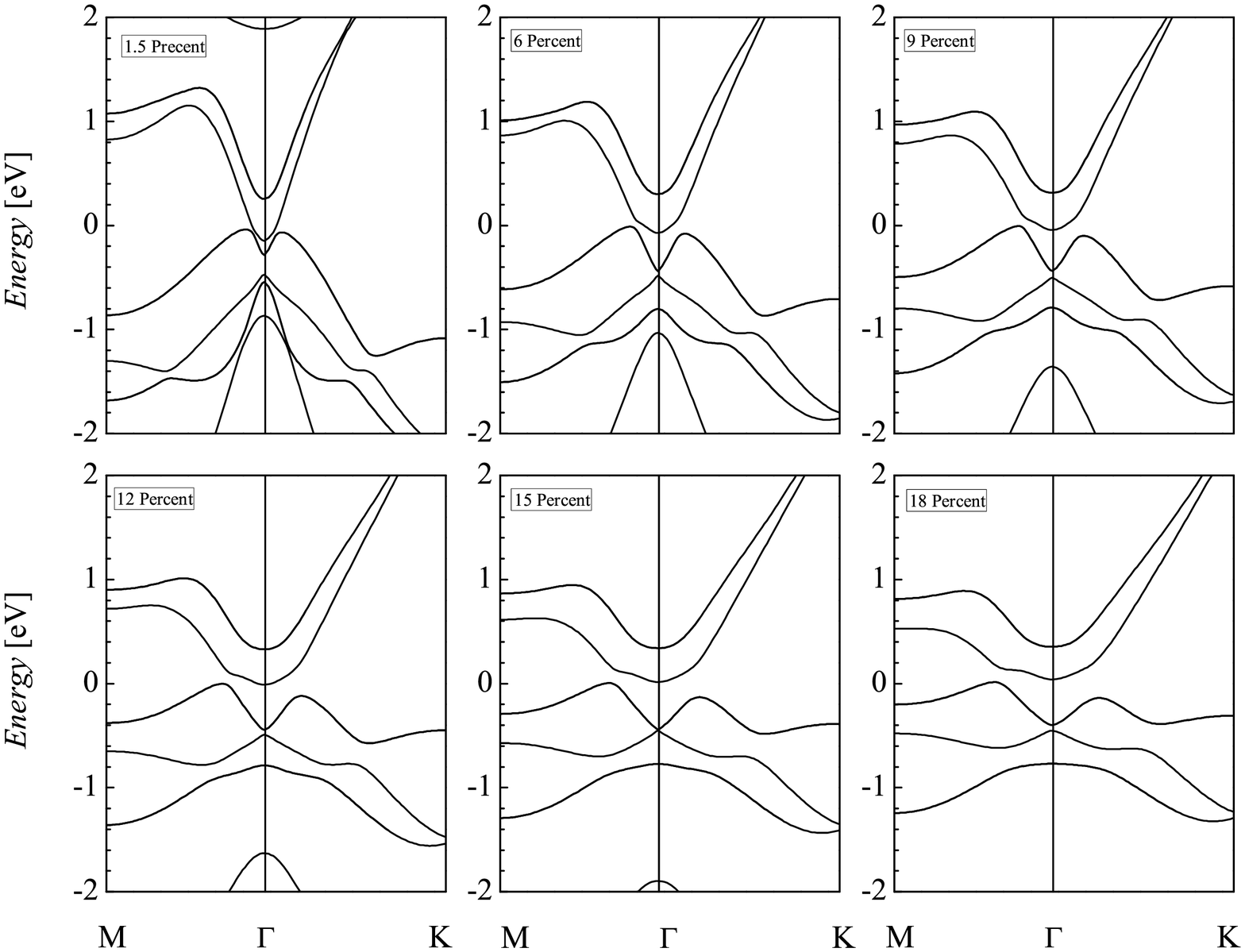}}
\caption{\label{Fig_6}The electronic band structure of FBSn in the DFT-D approximation
with tensile strain up to $\Delta a/a_{\circ}=0.18$.}
\end{figure}

We apply tensile strain to the hexagonal structure in order to open an energy band gap at the $\Gamma$ point. Figure \ref{Fig_6} shows the band structure of bilayer FBSn in the DFT-D with spin-orbit interaction under strain up to 0.18.\par By applying external strain the atomic bonds grow weaker which leads to reduction in electronic band widths. The applied strain reduces the band curvature and electronic velocities around the Fermi level. The electronic gaps at the $M$ and $K$ points are decreased but the main effect of tensile strain is for electronic bands around the $\Gamma$ point. The overlap between valence band maximum (VBM) and conduction band minimum (CBM) at the $\Gamma$ point is decreased and finally a very small electronic indirect electronic gap is opened at the $\Gamma$ point. To investigate the topological phase of FBSn we calculate the $\mathbb Z_2$ topological invariant by crossing the parities of occupied bands \cite{CWZhang2016, CWZhang2016_2, RWZhang2016, YWang2016}. Table \ref{Table} indicates the occupied bands paritiy at four TRIM points for FBSn unders 2 percent of biaxial tensile strain. For higher values of applied biaxial strain the $\mathbb Z_2$ topological invariant is same. The parity analysis reveals that even for relatively high values of strain, the band inversion does not occur in the FBSn structure and topological invariant is zero which indicates the trivial phase in strained structures. Although the monolayer fluorinated stanene has a spin-orbit gap of $0.3 eV$ \cite{Xu2013} and is an appropriate candidate for topological insulator, the bilayer stanene is semimetal and gapless. The applied tensile strain opens an indirect band gap around the Г point but the strained structure is a trivial insulator which is a drawback for application of fluorinated stanene as a topological insulator in electronic devices.

\begin{table}[!h]
\caption{\label{Table}$\mathbb Z_2$ Topological invariant}
\begin{center}
\begin{tabular}{ l c c }
\hline
$\Gamma_i$ & Parity of occupied bands & $\sigma_i$ \\
\hline
\hline
$\Gamma$  ~~ (0.0, 0.0) & +~-~+~-~+~-~-~+~+~-~+~-~-~+~-~+~+~-~+~-~+~-~+~-~-~-~+~+ & + \\
$M_1$ (0.5, 0.0) & +~-~+~-~+~-~+~-~+~-~+~-~+~-~+~-~+~-~-~+~+~-~+~-~+~-~+~- & + \\
$M_2$ (0.0, 0.5) & +~-~+~-~+~-~+~-~+~-~+~-~+~-~+~-~+~-~-~+~+~-~+~-~+~-~+~- & + \\
$M_3$ (0.5, 0.5) & +~-~+~-~-~+~+~-~+~-~+~-~+~-~+~-~+~-~-~+~+~-~+~-~+~-~+~- & + \\
\hline
Strained FBiSn & $\mathbb Z_2$ Topological invariant & $\mathbb Z_2$=1\\
\hline
\end{tabular}
\end{center}
\end{table}

\section{Conclusion}
In summary, we have studied the fluorinated bilayer stanene in the DFT model and consider the intrinsic spin-orbit and van der Waals interaction between different layers. Although the fluorinated monolayer stanene was predicted as a room temperature topological insulator\cite{Xu2013}, the bilayer stanene and fluorinated bilayer stanene are semi-metal. The the high values of applied tensile strain causes an electronic band gap and trivial insulator phase in the fluorinated bilayer stanene. 

\section{Acknowledgment}
One of the authors (Y. Mogulkoc) acknowledges “The Scientific and Technological Research Council of Turkey (TUBITAK)” through “BIDEB-2219 Postdoctoral Research Fellowship”. The numerical calculations reported in this paper were partially performed at TUBITAK ULAKBIM, High Performance and Grid Computing Center (TRUBA resources).


\section*{References}


\begin{thebibliography}{}

\bibitem{Kane2005}
C. L. Kane and E. J. Mele, Physical Review Letters {\bf 95} (2005) 146802.

\bibitem{CWZhang2012}
F. Zhang, and C. Zhang, Nanoscale Research Letters {\bf 4} (2012) 422.

\bibitem{CWZhang2012_2}
C. Zhang, and S Yan, The Journal of Physical Chemistry C {\bf 116} (2012) 4163.

\bibitem{CWZhang2016}
R. Zhang, C. Zhang, W. Ji, S. Li, S. Yan, S. Hu, P. Li, P. Wang, and F. Li, Scientific Reports {\bf 6} (2016) 18879.

\bibitem{CWZhang2016_2}
H. Zhao, C. Zhang, R. Zhang, S. Li, S. Yan, B. Zhang, P. Li, and P. Wang, Scientific Reports {\bf 6} (2016) 20152.

\bibitem{Hasan2010}
MZ. Hasan, CL. Kane,  Review Modern Physics {\bf 82} (2010) 3045.

\bibitem{Qi2011}
Q. L. Qi, S. C. Zhang, Review Modern Physics {\bf 83} (2011) 1057.

\bibitem{Yan2012}
B. Yan,S. C. Zhang, Reports on Progress in Physics {\bf 75} (2012) 096501.

\bibitem{Moore2013}
J. E. Moore, Nature Nanotechnology {\bf 8} (2013) 194.

\bibitem{Rasche2013}
B. Rasche, A. Isaeva, M. Ruck et al,  Nature Material {\bf 12} (2013) 422.

\bibitem{Nigam2015} 
S. Nigam, S. Gupta, D. Banyai, R. Pandey, and C. Majumder, Physical Chemistry Chemical Physics {\bf 17} (2015) 6705.

\bibitem{König2007}
M. König, S. Wiedmann, C. Brüne, A. Roth, H. Buhmann, L. W. Molenkamp, X.-L. Qi,
and S.-C. Zhang, Science {\bf 318} (2007) 766.

\bibitem{Hsieh2009}
D. Hsieh, Y. Xia, D. Qian, L. Wray, J. Dil, F. Meier, J. Osterwalder, L. Patthey, J.
Checkelsky, and N. Ong, Nature {\bf 460} (2009) 1101.

\bibitem{Tanaka2012}
Y. Tanaka, Z. Ren, T. Sato, K. Nakayama, S. Souma, T. Takahashi, K. Segawa, and Y.
Ando, Nature Physics {\bf 8} (2012) 800.

\bibitem{Chen2009}
Y. Chen, J. Analytis, J.-H. Chu, Z. Liu, S.-K. Mo, X.-L. Qi, H. Zhang, D. Lu, X. Dai, and Z.
Fang, Science {\bf 325} (2009) 178.

\bibitem{Kuroda2012}
K. Kuroda, M. Ye, A. Kimura, S. Eremeev, E. Krasovskii, E. Chulkov, Y. Ueda, K.
Miyamoto, T. Okuda, and K. Shimada, Physical Review Letters {\bf 105} (2010) 146801.

\bibitem{Chou2014} 
B.-H. Chou, Z.-Q. Huang, C.-H. Hsu, F.-C. Chuang, Y.-T. Liu, H. Lin, and A. Bansil, New Journal of Physics {\bf 16} (2014) 115008.

\bibitem{Cao2015}
G. Cao, Y. Zhang, and J. Cao, Physics Letters A {\bf 379} (2015) 1475.

\bibitem{R.Zhang2015}
R.-W. Zhang, C.-W. Zhang, W.-X. Ji, S.-S. Li, S.-J. Hu, S.-S. Yan, P. Li, P.-J. Wang, and
F. Li, New Journal of Physics {\bf 17} (2015) 083036.

\bibitem{Ren2015} 
Y. Ren, Z. Qiao, and Q. Niu, arXiv preprint arXiv:1509.09016 (2015).

\bibitem{Liu2011}
C.-C. Liu, W. Feng, and Y. Yao, Physical Review Letters {\bf 107} (2011) 076802.

\bibitem{H.Zhang2015}
H. Zhang, Y. Ma, and Z. Chen, Nanoscale (2015).

\bibitem{Wang2012}
X.-F. Wang, Y. Hu, and H. Guo, Physical Review B {\bf 85} (2012) 241402.

\bibitem{Xu2013}
Y. Xu, B. Yan, H.-J. Zhang, J. Wang, G. Xu, P. Tang, W. Duan, and S.-C. Zhang,
Physical Review Letters {\bf 111} (2013) 136804.

\bibitem{MoMo2016}
M. Modarresi, W. B. Kuang, T. P. Kaloni, M. R. Roknabadi, and G. Schreckenbach, AIP Advances {\bf 6} (2016) 095019.

\bibitem{CCLiu2011}
C.-C. Liu, H. Jiang, Y. Yao, Physical Review B {\bf 84} (2011) 195430.

\bibitem{Zhu2015}
F.-f. Zhu, W.-j. Chen, Y. Xu, C.-l. Gao, D.-d. Guan, C.-h. Liu, D. Qian, S.-C. Zhang, and
J.-f. Jia, Nature Materials {\bf 14} (2015) 1020.

\bibitem{Modarresi2015}
M. Modarresi, A. Kakoee, Y. Mogulkoc, and M. Roknabadi, Computational Materials
Science {\bf 101} (2015) 164.

\bibitem{Giannozzi2009}
P. Giannozzi, S. Baroni, N. Bonini, M. Calandra, R. Car, C. Cavazzoni, D. Ceresoli, G.
L. Chiarotti, M. Cococcioni, and I. Dabo, Journal of Physics: Condensed Matter {\bf 21}
(2009) 395502.

\bibitem{Perdew2010}
J. P. Perdew, K. Burke, and M. Ernzerhof, Physical review letters 77 (1996) 3865.
S. Grimme, J. Antony, S. Ehrlich, and H. Krieg, The Journal of Chemical Physics {\bf 132}
(2010) 154104.

\bibitem{Grimme2010}
S. Grimme, J. Antony, S. Ehrlich, and H. Krieg, The Journal of Chemical Physics {\bf 132} (2010) 154104.

\bibitem{Grimme2006}
S. Grimme, Journal of Computational Chemistry {\bf 27} (2006) 1787.

\bibitem{Monkhorst1976}
H. J. Monkhorst and J. D. Pack, Physical Review B {\bf 13} (1976) 5188.

\bibitem{Fu2007}
L. Fu and C. L. Kane, Physical Review B {\bf 76} (2007) 045302.

\bibitem{RWZhang2016}
R. Zhang, C. Zhang, W. Ji, P. Li, P. Wang, S. Li, and S. Yan, Applied Physics Letters {\bf 109} (2016) 182109.

\bibitem{YWang2016}
Y. Wang, W. Ji, C. Zhang, P. Li, F. Li, P. Wang, S. Li, and S. Yan, Applied Physics Letters {\bf 108} (2016) 073104,

\bibitem{Kokalj1999}
A. Kokalj, Journal of Molecular Graphics and Modelling {\bf 17} (1999) 176.

\bibitem{SZaveh2016}
S. J. Zaveh and M. R. Roknabadi and T. Morshedloo and M. Modarresi, Superlattices and Microstructures {\bf 91} (2016) 383.

\bibitem{Mirnezhad2012}
M. Mirnezhad, M. Modarresi, R. Ansari, and M. Roknabadi, Journal of Thermal
Stresses {\bf 35} (2012) 913.

\bibitem{Peng2013}
Q. Peng, C. Liang, W. Ji, and S. De, Physical Chemistry Chemical Physics {\bf 15} (2013)
2003.


\end{thebibliography}
\end{document}